\documentclass[aps,a4paper,superscriptaddress,nofootinbib,tightenlines,floats]{revtex4}
\usepackage{latexsym}
\usepackage{amssymb} 
\usepackage[dvips]{graphicx}
\begin{document}
\renewcommand{\topfraction}{1}
\renewcommand{\=}{\!\!=\!\!}

\title{F-term strings in the Bogomol'nyi limit are also BPS states}
\author{A. Ach\'{u}carro} \affiliation{Lorentz Institute of
Theoretical Physics, University of Leiden, 2333 RA Leiden, The
Netherlands} \affiliation{Department of Theoretical Physics, UPV-EHU,
Bilbao, Spain} \author{J. Urrestilla} \affiliation{Department of
Physics and Astronomy, University of Sussex, Brighton, U.K.}

\date{\today}

\begin{abstract}
We derive the Bogomol'nyi equations for supersymmetric Abelian F-term
cosmic strings in four-dimensional flat space and show that, contrary
to recent statements in the literature, they are BPS states in the
Bogomol'nyi limit, but the partial breaking of supersymmetry is from
N=2. The second supersymmetry is not obvious in the N=1 formalism, so
we give it explicitly in components and in terms of a different set of
N=1 chiral superfields. We also discuss the appearance of a second
supersymmetry in D-term models, and the relation to N=2 F-term
models. The analysis sheds light on an apparent paradox raised by the
recent observation that D-term strings remain BPS when coupled to N=1
supergravity, whereas F-term strings break the supersymmetry
completely, even in the Bogomol'nyi limit. Finally, we comment on their
semilocal extensions and their relevance to cosmology.
\end{abstract}

\maketitle

\def\rd{{\rm{d}}}
\def\d{\partial}
\def\half{{1 \over 2}}
\def\quarter{{1 \over 4}}
\def\simleq{\; \raise0.3ex\hbox{$<$\kern-0.75em
      \raise-1.1ex\hbox{$\sim$}}\; }
\def\simgeq{\; \raise0.3ex\hbox{$>$\kern-0.75em
      \raise-1.1ex\hbox{$\sim$}}\; }  
\def\L{{\mathcal L}}
\def\t{\tilde}      
\def\bea{\begin{eqnarray}}
\def\eea{\end{eqnarray}}
\def\be{\begin{eqnarray}}
\def\ee{\end{eqnarray}}
\def\eps{\epsilon}
\def\epsbar{\bar\epsilon}
\def\one{{(1)}}
\def\two{{(2)}}

\section{Introduction}

For a number of years now there has been considerable interest in the
cosmological impact of topological defects from supersymmetric
theories, in particular cosmic strings. Until now most of the efforts
had to do with identifying the cause of inflation.  $N\=1$ supersymmetric
models typically contain complex scalar fields, often with flat
directions in their potential, and so are natural candidates for
inflationary models. But they are also natural candidates to form
topological defects at the end of inflation. Cosmic strings in
supersymmetric theories can exhibit interesting new properties with
respect to their non-supersymmetric counterparts. In particular, both
the fermionic zero modes arising from partial supersymmetry breaking
and the bosonic zero modes coming from flat directions or Goldstone
bosons can have a very serious impact on the cosmological evolution of
such defects.

More recently, interest was further revived by the realization that
the soliton solutions in four-dimensional supergravity might be
directly identified with ten-dimensional superstring states whose low
energy limit they represent \cite{DKP04,CMP04,DV04}. If these supergravity
defects have a potentially measurable effect on the early Universe,
comparison with cosmological data might provide direct information
about superstring/M theory. One idea that is central to this
philosophy is that the BPS states in a supersymmetric model will
survive all sorts of deformations of the model. They provide the
necessary connection with the higher energy theory.
  
It is well known that solitons whose mass saturates a Bogomol'nyi-type
bound generically lead to partial breaking of supersymmetry, and this
in turn protects the energy bound from receiving quantum
corrections. States which preserve only a subset of the supersymmetry
present in the vacuum are known are BPS states (after
Bogomol'nyi-Prasad-Sommerfield monopoles) and they are global minima
of the energy in their topological sector. Although these two
conditions are different, they are connected by the appearance of
central charges in the supersymmetry algebra \cite{WO78,HP86,
HS92a,HS92b,HS93}. Bogomol'nyi bounds usually survive
the coupling to gravity \cite{CG87}, and we are not aware of
any examples where coupling to gravity destroyed the BPS nature of the
states.

Dvali, Kallosh and van Proeyen \cite{DKP04} addressed the interesting
question of what supersymmetric BPS vortex solutions are possible in
$N\=1$ supergravity. In the absence of gravity, this question was
considered by Davis, Davis and Trodden \cite{DDT97} some years ago.
There are two known ways to spontaneously break a gauge symmetry in
$N\=1$ (globally) supersymmetric theories: either by using Fayet-Iliopoulos 
D-terms (which only works in the Abelian case) or by adding
F-terms to the superpotential. The Higgs mechanism then leads to the
formation of Nielsen-Olesen strings which are usually known as D-term
strings and F-term strings respectively.
The existence of these string solutions in $N\=1$ supergravity was
established by Morris \cite{M97}, although he did not consider the
supersymmetry breaking.  In 2+1 dimensions the problem was analysed in
\cite{BBS95,ENS96}.

Davis, Davis and Trodden concluded that (global, $N\=1$) supersymmetry
was partially broken by D-term strings and fully broken by F-term
strings, even in the Bogomol'nyi limit. Dvali, Kallosh and van Proeyen
conclude that this result is still true when the models are coupled to
$N\=1$ supergravity.

Their results would seem to put into question the BPS nature of the
F-strings in the Bogomol'nyi limit. In fact, the issue is more
subtle. As one might expect on general grounds
\cite{HS92a,HS92b,HS93,ENS94}, the globally supersymmetric model
has a second supersymmetry in the Bogomol'nyi limit which is not
evident in $N\=1$ language. As we will show, before coupling to gravity
both D- and F-term strings are BPS states {\it that preserve 1/2 of
the $N\=2$ supersymmetry} \cite{VY01}. It is therefore remarkable that in $N\=1$
supergravity D-term strings remain BPS states, whereas F-term strings
break all (local) supersymmetries {\it even in the Bogomol'nyi limit}
\cite{DKP04}.

While this result appears paradoxical, we should quickly point out
that there is no contradiction, in principle. First of all, the global
limit of $N\=2$ local supersymmetry is not at all trivial.
Secondly, the Higgs mechanism that gives rise to the strings is due to
constant Fayet-Iliopoulos terms, which are known to be incompatible
with $N\=2$ supergravity (see \cite{K01} for references). Although the
Bogomol'nyi bound survives coupling to bosonic gravity, the coupling to
$N\=1$ supergravity breaks the $SU(2)$ global symmetry that relates F-
and D-term models in the global case and so we should not expect
F-term and D-term models to remain equivalent once they are coupled to
$N\=1$ supergravity. We will get back to this point at the end of the
paper.

In D-term models the second supersymmetry only appears when a specific
trilinear coupling is present in the superpotential. The trilinear
coupling has to satisfy a Bogomol'nyi-looking relation that we call the
superBogomol'nyi limit\footnote{The term superBogomol'nyi is sometimes
used in the literature to refer to the ordinary Bogomol'nyi limit in
F-term models; here we mean something different} (because the ordinary
Bogomol'nyi limit is always satisfied in D-term models).

The purpose of this paper is to take over where \cite{DDT97} left off and
analyse the effect of the second supersymmetry in both F-term models
in the Bogomol'nyi limit and D-term models in the superBogomol'nyi
limit. We have made a point of avoiding the $N\=2$ formalism, which is
not so familiar to cosmologists\footnote{See \cite{ARH98,ADPU02,KL03} and \cite{VY01}
for a full N=2 treatment}. In some sense this is an attempt to
translate the $N\=2$ results into $N\=1$ language. Since the second
supersymmetry has the opposite chirality, we make it explicit in a
(non-standard) way that makes partial breaking of supersymmetry easier
to study.

\section{Before supersymmetry}

We begin by considering the bosonic Abelian Higgs model, with action
\be 
S =\!\! \int d^4x\left[|D_\mu \phi|^2 - \frac1 4{F_{\mu\nu}
F^{\mu\nu}}
- {\frac\lambda 2}(|\phi|^2 - \eta^2)^2 \right] 
\label{AH}
\ee 
where $A_\mu$ is a $U(1)$ gauge field and $\phi$ is a complex
scalar of charge $e$ ,
\be
D_{\mu}\phi  = (\partial_{\mu} +  i \, e\, A_{\mu})\,\phi,
 \qquad
F_{\mu\nu} = \partial_\mu A_\nu - \partial_\nu A_\mu
\ee
We want to study the properties of straight, static vortices along the
$z$-direction so we drop the $t$- and $z$-dependence and set~$A
_t$~$=$~$A_z$~$=$~$0$.  Their energy per unit length is
\begin{equation}
E =\!\! \int d^2x\left[|D_1\phi|^2 + |D_2\phi|^2 + \frac1
2 B^2 
+{\frac\lambda 2}(|\phi|^2 - \eta^2
)^2 \right]
\end{equation}
where $B = \partial_1 A_2 - \partial_2 A_1$ is (the $z$-component of)
the magnetic field. Finite energy configurations must satisfy $|\phi|
\to \eta$ as $r \to \infty$ (the vacuum manifold is a circle)
but also $\ D_\mu \phi \to 0 , B \to 0 $ faster than $1/r$. If we
choose the gauge $A_r = 0$, cylindrically symmetric configurations
tend to $\phi (r, \theta) \sim \eta e^{i n \theta}$, $ A_\theta (r,
\theta) \sim -n / (er) $, as $r \to \infty$, and this means that the
total magnetic flux in the plane perpendicular to the string is
quantized,
\be
\int d^2x B = \oint {\vec A}\cdot {\vec dl} = - {2\pi n \over e}
\ee
Following Bogomol'nyi we use the identity $ [D_1, D_2] \phi = i e B \phi
$ and an integration by parts  to rewrite the energy as
\begin{eqnarray}
E =\!\! \int d^2x ~ &\left[| (D_1  \pm i D_2)\phi|^2
+{1 \over 2} [B \mp e (|\phi|^2 - \eta^2)]^2
+{\lambda - e^2 \over 2}(|\phi|^2 - \eta^2
)^2 \right] \cr
&\mp  \eta^2 e \int d^2 x ~B
\end{eqnarray}
We have omitted a boundary term, the curl of a current ${\vec J} = -i
{\phi^*} {\vec D}\phi$ which vanishes at $r = \infty$. The last term
is the topological charge $\mp \eta^2 e \int B = \pm 2\pi n \eta^2 =
T$. If $\lambda = e^2$, the energy has a lower bound $E \geq |T|$, and
the minimum energy configurations ($E = |T|$) are those that satisfy
\begin{eqnarray}
(D_1 + i D_2 ) \phi &=& 0 , \qquad B - e(\phi^2 - \eta^2) = 0 \ \qquad {\rm if}
\ \ n >0 \cr
(D_1 - i D_2 ) \phi &=& 0 , \qquad B +  e(\phi^2 - \eta^2) = 0 \ \qquad 
{\rm if} \ \ n < 0
\label{bogovortexeqn}
\end{eqnarray}
The condition $\lambda = e^2$ is known as the Bogomol'nyi limit.  In
physical terms, it states that the mass of the scalar fluctuations
$m_s = \sqrt {2\lambda} \eta$ is equal to the mass of the vector
fluctuations $m_v = \sqrt 2 e\eta$. Perhaps for this reason the
Bogomol'nyi limit is sometimes called the supersymmetric limit,
although this is somewhat misleading because it is not a necessary
condition for $N\=1$ supersymmetrization\footnote{Actually, the charged
scalar and the gauge field must always belong to different
supermultiplets.}  (see the F-term models below). 

Notice that if $\lambda < e^2$ (or $m_s < m_v$) the last term in the
 integral is not positive, so we cannot deduce a lower bound on the
 energy. If $\lambda > e^2$ ($m_s >m_v$) we have a lower bound, $E\geq
 |T|$, but it can never be attained by a vortex configuration since
 the conditions $B \pm e(\phi^2 - \eta^2) = 0$ and $(\phi^2 - \eta^2)
 = 0$ taken together imply $B=0$, which is incompatible with a total
 magnetic flux of $2\pi n /e$ for non-zero $n$ ~($n=0$ gives
 $\phi=\pm\eta$, the vacuum).

We can take $n>0$ without loss of generality. From now on we will
restrict ourselves to the lowest non-trivial topological sector,
$n=1$.  If we consider cylindrically symmetric configurations, 
\be
\phi
= \eta f(r) e^{i\theta} \ \qquad\qquad A_\theta = -(1 / e) {a(r) \over
r}
\label{NOprofiles} 
\ee 
with $f(0)=a(0)=0$, $f(\infty) =
a(\infty) = 1$, the equations (\ref{bogovortexeqn}) become
\begin{equation} 
f' + {a-1 \over r} f = 0  \qquad {a'\over r} + e^2 \eta^2(f^2 - 1) = 0
\label{NOeqs}\end{equation} 
known as the Nielsen--Olesen equations.

In what follows we will consider two supersymmetrizations of the
Abelian Higgs model which allow Nielsen--Olesen vortices as stable
solutions. The names
``F-term'' and ``D-term'' come from $N\=1$ supersymmetry, where they
refer to the F- and D- auxiliary fields of the $N\=1$ chiral and gauge
superfields, respectively. 

Before we go into the supersymmetrizations of the Abelian Higgs model,
we should mention the semilocal model, which is obtained when the
charged field is not a complex scalar but an SU(2) doublet $(\phi_1,
\phi_2)$ \cite{AV00}. The presence of Goldstone bosons changes the
stability properties of the strings in a dramatic way.  Although
magnetic flux is still quantized on all solutions with finite energy
per unit length, there is no guarantee that the flux will be confined
to a core of a finite size. String configurations are only stable if
$m_s/m_v <1$. In the Bogomol'nyi limit the string is
only neutrally stable and now there is a zero mode which makes the
magnetic flux spread to an arbitrarily large area. This zero model is
easily excited in a cosmological context, so one does not expect
stringy configurations to appear in a cosmological phase
transition. It turns out that supersymmetric versions of the semilocal
model appear naturally in some brane inflation scenarios and other
superstring-inspired models \cite{UAD04,BDKP04,DHKLZ04}.

\section{$N\=1$ supersymmetric Abelian Higgs models}

In the next three sections we review and expand the analysis of
\cite{DDT97} to F-term models in the Bogomol'nyi limit and D-term
models in the superBogomol'nyi ($N\=2$) limit.  Unless otherwise stated,
we follow the conventions of \cite{DDT97}.  In particular, from now on
$e = g/2$.

Let us consider supersymmetric QED, containing an Abelian vector
superfield $V$, and $m$ chiral superfields $\Phi_i$, $(i=1,...,m)$
with $U(1)$ charges $q_i$. In order to have a theory without anomalies
which does not leave the gauge symmetry unbroken, we need at least 3
different chiral superfields $\Phi_0,\,\Phi_+,\,\Phi_-$ with charges
$0,+1,-1$ respectively. The detailed field content of these models is
the following: two charged chiral superfields $\Phi_\pm =
(\phi_\pm, \psi_\pm, F_\pm)$, one neutral chiral superfield $\Phi_0 =
(\phi_0, \psi_0, F)$ and the vector superfield $V=(A_\mu,\lambda,D)$
in the Wess-Zumino gauge. The fields 
$\phi_\pm$ and $\phi_0$ are complex scalar
fields; $A_\mu$ is a $U(1)$ gauge field; $\psi_\pm$, $\psi_0$ and
$\lambda$ are Weyl fermions; and $F_\pm$, $F$ and $D$ are auxiliary
fields.

We can also have a superpotential, the most general form for a
renormalisable theory being \be W(\Phi_i)=a_i\Phi_i+\half
b_{ij}\Phi_i\Phi_j+\frac{1}{3}\Phi_i\Phi_j\Phi_k \ee As each term in
the superpotential has to be gauge invariant, $a_i\ne0$ only if
$q_i=0$, $b_{ij}\ne0$ only if $q_i+q_j=0$ and $c_{ijk}\ne0$ only if
$q_i+q_j+q_k=0$.  When the model is Abelian (which is our case) we can
also add a Fayet-Iliopoulos term of the form $\xi_3 D$ (the reason for
the subindex in $\xi_3$ will become clear later).

The Lagrangian density for these models in Wess-Zumino gauge is given
by \cite{DDT97} \be \L=\L_B+\L_F+\L_Y-U \ee with \bea
\L_B&=&\left|D_\mu\phi_i\right|^2-\quarter\,F^{\mu\nu}F_{\mu\nu}\nonumber\\
\L_F&=&-i\psi_i\sigma^\mu
D_\mu\bar\psi_i-i\,\lambda_i\sigma^\mu\d_\mu\bar\lambda_i\nonumber\\
\L_Y&=&i\frac{g}{\sqrt{2}}q_i\phi_i^*\psi_i\lambda-\left(b_{ij}+c_{ijk}\phi_k\right)\psi_i\psi_j+({\rm
h.c})\nonumber\\
 U&=&|F_i|^2+\half
D^2=|a_i+b_{ij}\phi_j+c_{ijk}\phi_i\phi_j\phi_k|^2+\half\left(\xi_3+\frac{g}{2}q_i\phi_i^*\phi_i\right)^2
\label{lagr}
\eea where $D_\mu \phi_i=(\d_\mu+ \half i\,g q_i A_\mu)\phi_i$ and we
have substituted the equations of motion of the auxiliary fields into
the Lagrangian.

\section
{D-term strings}

As discussed in \cite{DDT97}, the simplest model containing D-term
strings has a Lagrangian 
(\ref{lagr}), with a
Fayet-Iliopoulos term $\xi_3=-\half g \eta^2$ 
and no
superpotential. The vacuum manifold in this case is given by \be
U=\frac{g^2}{8}\left(|\phi_+|^2-|\phi_-|^2-\eta^2\right)^2
\label{vacd}
\ee

We are interested in a static, straight string with winding number $n=1$
in the $z$-direction. The energy per unit length of such an object
satisfies a Bogomol'nyi bound:

\begin{eqnarray}
E&=&\int\!\!\rd^2x\left[|D_1\phi_+|^2+|D_2\phi_+|^2+|D_1\phi_-|^2+|D_2\phi_-|^2+
\frac{g^2}{8} \left(|\phi_+|^2-|\phi_-|^2-\eta^2\right)^2+\half
B^2\right]\nonumber\\ &=&\int\!\!\rd^2x\left\{\left|\left(D_1+i
D_2\right)\phi_+\right|^2+\left|\left(D_1+ i D_2\right)\phi_-\right|^2
+ \half
\left[B-\frac{g}{2}\left(|\phi_+|^2-|\phi_-|^2-\eta^2\right)\right]^2\right\}
+ T
\label{ener}
\end{eqnarray}
where, as in the bosonic case, \be T = - \frac{g}{2}\eta^2\int\!\!\rd^2x
B = \frac{g}{2}\eta^2 2\pi n \ee is the topological charge (in the
transverse plane). 

We can read off the Bogomol'nyi equations directly from (\ref{ener})
\bea &(D_1+iD_2)\phi_+=0\,,\qquad(D_1+iD_2)\phi_-=0&\nonumber\\
&B-\frac{g}{2}\left(|\phi_+|^2-|\phi_-|^2-\eta^2\right)=0& \ \  .
\label{bogol}
\eea
Even though the vacuum manifold (\ref{vacd}) includes configurations
with $\phi_- \neq 0$, only the $\phi_-=0$ configurations saturate the
Bogomol'nyi bound \cite{ADPU02}.  The remaining fields $(\phi_+, \
A_\mu)$ take the usual Nielsen-Olesen vortex form
(\ref{NOprofiles},\ref{NOeqs}).

The condition $\phi_- = 0$ is a direct consequence of the first two
Bogomol'nyi (\ref{bogol}) equations, which imply $(\partial_1 +
i\partial_2)(\phi_+ \phi_-) = 0$ or, using $z=x+iy$,
$\d_z(\phi_+\phi_-)=0$. Since $(\phi_+\phi_-)$ is analytic and bounded
for all z, it is constant.  For a string, with $\phi_+(0) = 0$, the
constant must be zero, and so we conclude that $\phi_-\equiv 0$
\cite{ADPU02,PRTT96,PU03}.

Note that we did not have to impose any condition on the parameters of
the model, the Bogomol'nyi condition $\lambda = g^2/4$ in (\ref{AH}) is
automatically satisfied.  The string solution saturates the
Bogomol'nyi bound, and so we expect partial breaking of
supersymmetry. This is easily verified by looking at the supersymmetry
transformations of the fermions in the background given by the string
(\ref{bogol}) with all fermions put to zero \cite{DDT97} (the bosonic
fields are obviously invariant since their transformations are
proportional to the fermions) \bea & &\delta (\psi_+)_\alpha = i
\sqrt{2}
\left(\sigma^1D_1\phi_++\sigma^2D_2\phi_+\right)_{\!\!\alpha\dot\alpha}\epsbar^{\dot\alpha\one}=2iD_1\phi_+\left(\begin{array}{cc}
0 & 1 \\ 0 &
0\end{array}\right)_{\alpha\dot\alpha}\epsbar^{\dot\alpha\one}
\nonumber\\ & &\delta (\psi_-)_\alpha = 0\nonumber\\ &
&\delta(\psi_0)_\alpha = 0 \nonumber\\ & &\delta (\lambda)_\alpha =
i\left(\sigma^3 B +
\frac{g}{2}(|\phi_+|^2-\eta^2)\right)_{\!\!\alpha}^{\!\!\beta}\eps^\one_\beta
=2iB \left(\begin{array}{cc}1 & 0 \\ 0 &
0\end{array}\right)_{\!\!\alpha}^{\!\!\beta}\eps^\one_\beta \eea where
$\eps^\one$ is the SUSY parameter.

The string solution is invariant under half of the supersymmetry
transformations: those with $\eps^\one_1=0$.
The other half are broken,
they create fermion zero modes. The number of fermionic zero modes is
half what one would expect in an arbitrary configuration with positive
energy.  We say that the string is a $\half$-BPS state, or that the
supersymmetry is half-broken.

As mentioned in the introduction, at a given limit the theory becomes $N\=2$ supersymmetric.
We wish to find the second supersymmetry from an $N\=1$ point of view: a reordering of
the fields  in new multiplets will allow us to show the second supersymmetry explicitly. First, for reasons that will become clear later, we have to add a superpotential to the model:
\be W=\beta\Phi_0\Phi_+\Phi_-
\label{supd}
\ee 
which modifies the scalar potential to 
\be
U=\frac{g^2}{8}\left(|\phi_+|^2-|\phi_-|^2-\eta^2\right)^2+\beta^2|\phi_0|^2\left(|\phi_-|^2+|\phi_+|^2\right)+\beta^2|\phi_+\phi_-|^2
\label{vac2}
\ee 
The new
vacuum manifold includes the condition $\phi_-=\phi_0=0$
automatically, without any further vacuum selection effect by the
strings.

The Yukawa terms after the addition of the superpotential 
(\ref{supd}) are the following:
\be
\L_Y=i\frac{g}{\sqrt{2}}\left(\phi_+^*\psi_+\lambda+\phi_-^*\psi_-\lambda\right)
-\beta\left(\phi_0\psi_+\psi_-+\phi_+\psi_-\psi_0+\phi_-\psi_+\psi_0\right)
+({\rm
h.c})
\label{yuk}
\eea

The new supersymmetry will re-shuffle the Yukawa terms. Therefore, in order to leave the Yukawa terms
invariant with respect to the new supersymmetry,
a  specific value for $\beta$ is needed
\cite{ENS96}.
We call this limit the superBogomol'nyi limit
\be
\beta=\frac{g}{\sqrt{2}}
\label{superb}
\ee

Given the D-term model (\ref{lagr}), with a
Fayet-Iliopoulos term $\xi_3=-\half g \eta^2$,superpotential (\ref{supd}) and 
in the superBogomol'nyi limit, we can try to
find the second supersymmetry from an $N\=1$ point of view:

The second supersymmetry is
anti-chiral, and connects the positively (negatively) charged scalar
to the charge conjugate of the negatively (positive) charged fermion,
and the neutral scalar to the gaugino. We can write it as chiral
provided we exchange the scalar particles and their charge conjugates
in the multiplets.

Consider the following multiplets: \bea &
&\hat\Phi_+=\left(\phi_-^*,\psi_+,\hat F_+\right)\nonumber\\ &
&\hat\Phi_-=\left(-\phi_+^*,\psi_-,\hat F_-\right)\nonumber\\ &
&\hat\Phi_0=\left(\phi_0,i\lambda,\hat F_0\right)\nonumber\\ & &\hat
V=\left(A_\mu,i\psi_0,\hat D\right)
\label{newdmultiplets}
\eea

This way of reordering the fields into new multiplets is already
 foreseeing that the neutral chiral multiplet ($\Phi_0$) and the gauge
 ($V$) will be part of the same $N\=2$ vector multiplet; and similarly
 with $\Phi_+$ and $\Phi_-$ being part of the same
 hypermultiplet. 

It is easy to check that from a field content given by the ``new`` $N\=1$
supersymmetric multiplets (\ref{newdmultiplets}), we can
obtain the D-term model provided that
we add
a superpotential of the form
$W=\beta\hat\Phi_0\hat\Phi_+\hat\Phi_-$ and a Fayet-Iliopoulos term
$\hat\xi_3=\half g \eta^2 = -\xi_3$ (note the change in sign in the Fayet-Iliopoulos 
term). The Yukawa terms are identical to
(\ref{yuk}) if we are in the superBogomol'nyi limit.
Therefore, we have been able to arrive at the same model from two different sets of 
supersymmetric multiplets, i.e., the model possesses two supersymmetries.

Under this new supersymmetry, we have new fermionic zero modes \bea &
& \delta (\psi_+)_\alpha =  0\nonumber\\
& &\delta (\psi_-)_\alpha =  -
i \sqrt{2}
\left(\sigma^1D_1\phi_+^*+\sigma^2D_2\phi_+^*\right)_{\alpha\dot\alpha}
\epsbar^{\dot\alpha\two}=-2iD_1\phi_+^*\left(\begin{array}{cc} 0 & 0
\\ 1 & 0\end{array}
\right)_{\alpha\dot\alpha}\epsbar^{\dot\alpha\two}\nonumber\\ 
& &\delta\psi_0 = \left(\sigma^3 B -
\frac{g}{2}(|\phi_+|^2-\eta^2|)\right)_\alpha^\beta\eps^\two_\beta=2B
\left(\begin{array}{cc}0 & 0 \\ 0 &
1\end{array}\right)_\alpha^\beta\eps^\two_\beta \nonumber\\
& &
\delta \lambda = 0
\eea

Therefore, we have 2 supersymmetries half broken, and a total of four
fermionic zero modes. The projected supersymmetry generators
$\sigma_+\eps^\one \,, \sigma_-\eps^\two$ are the generators of the
fermionic zero modes, whereas $\sigma_-\eps^\one \,,
\sigma_+\eps^\two$ generate the unbroken supersymmetries.  The
projectors $\sigma_\pm$ are given by: \be
\sigma_+=\half\left(1-i\sigma^1\sigma^2\right)=\left(\begin{array}{cc}
1 & 0 \\ 0 & 0\end{array}\right)\qquad\qquad
\sigma_-=\half\left(1+i\sigma^1\sigma^2\right)=\left(\begin{array}{cc}
0 & 0 \\ 0 & 1\end{array}\right)
\label{pro}
\ee

The D-term model can be made semilocal by the addition of a second
pair of charged superfields $\Psi_+,\Psi_-$. The superpotential \be
W=\beta\Phi_0(\Phi_+\Phi_- + \Psi_+\Psi_-) \ee gives a scalar
potential (note that $\psi_\pm$ are now scalars) 
\be
U=\frac{g^2}{8}\left(|\phi_+|^2 + |\psi_+|^2 -|\phi_-|^2
-|\psi_-|^2
-\eta^2\right)^2+\beta^2|\phi_0|^2\left(|\phi_-|^2+|\phi_+|^2 +
|\psi_-|^2 + |\psi_+|^2 \right)+\beta^2|\phi_+\phi_-+\psi_+\psi_-|^2 
\ee 
The Bogomol'nyi limit is automatically
satisfied for any $\beta$ and the configurations that saturate the
bound have $\phi_-=\phi_0=\psi_-=0$. However there is no unique string
solution, but a one-parameter family of string-like configurations
where the magnetic flux is spread over an arbitrarily large area.
The general solution, modulo $SU(2)$ rotations between $\phi_+$ and $\psi_+$,\footnote{Note that this $SU(2)$ rotation
is not related to the $N=2$ $SU(2)$} can be given - for suitable chosen
$f(r)$, $a(r)$ depending on $q_0$ - by \cite{H92}
\be
\phi_+=\eta f(r) e^{i\theta}\quad\psi_+= q_0 \frac{f (r )}{r}
 \quad\ 
A_\theta = -\frac{2}{ g} \frac{a(r)}{r}
\ee
All these configurations saturate the energy bound. The parameter $q_0$ fixes the size for the vortex: for $q_0=0$, we recover the Nielsen-Olesen profile (\ref{NOprofiles}), and for bigger $q_0$ the width of the resulting string is larger.
In fact, a small perturbation on any of the strings of the family will cause the string to become wider \cite{l92}. 
The magnetic flux remains quantized, although in the wider strings it is more
spread out (see also \cite{EY03}). 

Irrespective of the value of $q_0$, the configurations are all BPS. In the superBogomol'nyi limit the second
supersymmetry appears and the number of fermionic zero modes doubles \cite{ADPU03}.

\section
{F-term strings}\label{Ftermstrings}

F-term string solutions can be obtained with the following
superpotential \be W= \beta\Phi_0 (\Phi_+ \Phi_- - \frac{\eta^2}{2})
\label{supf}
\ee where the last term is known as an F-term. There is no need of
including a Fayet-Iliopoulos term.  The potential energy for the
present case reads: 
\be
U=\beta^2\left|\phi_+\phi_--\frac{\eta^2}{2}\right|^2+\beta^2|\phi_0|^2\left(|\phi_+|^2+|\phi_-|^2\right)+\frac{g^2}{8}\left(|\phi_+|^2-|\phi_-|^2\right)^2
\ee

If $\beta = g^2 /2$, we can write the energy per unit length of the
system in the Bogomol'nyi way, for a static, straight string with
winding number $n$ in the z-direction: \bea E&=& \int \rd^2 x
\half\left[\left|(D_1+iD_2)(\phi_++\phi_-^*)\right|^2+\half\left|(D_1+iD_2)(\phi_--\phi_+^*)\right|^2+
\half \left[B- \frac{g}{2}( \phi_+\phi_-+\phi_+^*\phi_-^* -
\eta^2)\right]^2+\right.\nonumber\\ &
&\frac{g^2}{8}\left(i\phi_+\phi_--i\phi^*_+\phi_-^*\right)^2 +
\frac{g^2}{8}\left(|\phi_+|^2 - |\phi_-|^2\right)^2
+\frac{g^2}{2}|\phi_0|^2\left(|\phi_+|^2+|\phi_-|^2\right) \bigg]
\nonumber\\ & &+\frac{g}{2}\eta^2\int \rd^2 x B
\label{bf}
\eea Note that, contrary to the D-term case, we had to impose the
condition (\ref{superb}) in order to be able to write the energy in
the form (\ref{bf}). This is the (ordinary) Bogomol'nyi limit.

The Bogomol'nyi equations can then be immediately read off
\bea 
& &(D_1+iD_2)(\phi_++\phi_-^*)=0\,; \qquad \qquad (D_1+iD_2)(\phi_--\phi_+^*)=0\,;
\label{Bogomol'nyi}\\ 
& &|\phi_+|^2-|\phi_-|^2=0\,; \qquad \qquad \phi_+\phi_-=\phi_+^*\phi_-^*\nonumber\\
& & B-\frac{g}{2}(
\phi_+\phi_-+\phi_+^*\phi_-^* - \eta^2) =0\,.
\nonumber 
\eea
which can be rewritten as
\bea
& & \phi_+=\phi_-^*\nonumber\\
& & (D_1+iD_2)\phi_+=0\nonumber\\
& & B-g\left(|\phi_+|^2-\frac{\eta^2}{2}\right)=0\nonumber\\
\eea
Bogomol'nyi equations have been derived before in the literature
\cite{H01,GS00} using the ansatz $\phi_+ = { \phi_-}^*
$. We just showed here that the condition $\phi_+ = {\phi_-}^*$ is
not an ansatz, but one of the Bogomol'nyi equations. 

From the point of view of $N\=1$ supersymmetry, the straight infinite
string configuration breaks supersymmetry completely. This can be seen
from the fermionic supersymmetry transformations: \bea & &\delta
(\psi_+)_\alpha = i \sqrt{2}
\left(\sigma^1D_1\phi_++\sigma^2D_2\phi_+\right)_{\!\!\alpha\dot\alpha}
\epsbar^{\dot\alpha\one} \nonumber\\ & &\delta (\psi_-)_\alpha = i
\sqrt{2}
\left(\sigma^1D_1\phi_-+\sigma^2D_2\phi_-\right)_{\!\!\alpha\dot\alpha}
\epsbar^{\dot\alpha\one}\nonumber\\ & &\delta(\psi_0)_\alpha =
-g\left(|\phi_+|^2-
\frac{\eta^2}{2}\right)_{\!\!\alpha}^{\!\!\beta}\eps^\one_\beta
\nonumber\\ & &\delta (\lambda)_\alpha = i B
\left(\sigma^3\right)_{\!\!\alpha}^{\!\!\beta}\eps^\one_\beta \eea

In the Bogomol'nyi limit, one would expect to have some unbroken
supersymmetry remaining, i.e., some of the fermionic transformations
should be zero. Applying the Bogomol'nyi equations to those fermionic
zero modes we get 
\bea & &\delta (\psi_+)_\alpha = i 2\sqrt{2}
D_1\phi_+\left(\begin{array}{cc} 0 & 1 \\ 0 & 0
\end{array}\right)_{\!\!\alpha\dot\alpha} \epsbar^{\dot\alpha\one}
\nonumber\\
& &\delta (\psi_-)_\alpha =  i 2\sqrt{2}
D_1\phi_-\left(\begin{array}{cc} 0 & 0 \\ 1 & 0\end{array}
\right)_{\!\!\alpha\dot\alpha} \epsbar^{\dot\alpha\one}\nonumber\\ 
& &\delta(\psi_0)_\alpha = - B \eps^\one_\alpha \nonumber\\
& &\delta (\lambda)_\alpha =  i B
\left(\sigma^3\right)_{\!\!\alpha}^{\!\!\beta}\eps^\one_\beta
\eea
Unlike in the D-term case, there is no way of finding a projection of
the SUSY parameter that makes all four fermions zero. In the
Bogomol'nyi limit some of the entries cancel but the cancellations
happen in such a way that there is some fermionic content left. So it
appears that there are twice as many fermion zero modes in F-strings
than in D-strings.

But in the Bogomol'nyi limit the system has $N\=2$ supersymmetry, and
it is closely related to the D-term model (in fact, they are
equivalent). So let us try to search for the second supersymmetry as
in the previous case: analysis of the Yukawa couplings
(similar to the D-term case
studied in the previous section) shows that with
no Fayet-Iliopoulos term ($\hat\xi_3=0$) and superpotential \be
W=\frac{g}{2}\hat\Phi_0\left(\hat\Phi_+\hat\Phi_-+\frac{\eta^2}{2}\right)
\ee we obtain the Lagrangian for F-term strings. The hatted multiplets
are the same as before, eq.  (\ref{newdmultiplets}). Note the change of sign
in the $\eta^2$ term in the superpotential. We can then read the new
supersymmetric transformations (in the Bogomol'nyi limit) 
\bea &
&\delta (\psi_+)_\alpha =  i 2\sqrt{2}
D_1\phi_+\left(\begin{array}{cc} 0 & 1 \\ 0 & 0
 \end{array}\right)_{\!\!\alpha\dot\alpha} \epsbar^{\dot\alpha\two}
\nonumber\\ & &\delta (\psi_-)_\alpha = -i 2\sqrt{2}
D_1\phi_-\left(\begin{array}{cc} 0 & 0 \\ 1 & 0\end{array}
\right)_{\!\!\alpha\dot\alpha} \epsbar^{\dot\alpha\two}\nonumber\\ &
&\delta(\psi_0)_\alpha = B
\left(\sigma^3\right)_{\!\!\alpha}^{\!\!\beta}\eps^\two_\beta
\nonumber\\ 
& &\delta (\lambda)_\alpha = -i B \eps^\two_\beta \eea
Once again, there is no unbroken supersymmetry, until we consider
combinations of these two supersymmetries: \be
\left(\delta^\one-\delta^\two\right)\psi_+ =  0 \qquad\qquad
\left(\delta^\one+\delta^\two\right)\psi_- =  0 \ee It can easily be
seen that
\be
\sigma_-\left(\delta^\one+\delta^\two\right)f=0\qquad\qquad
\sigma_+\left(\delta^\one-\delta^\two\right)f=0 \ee for $f$ any
fermion in the theory. Thus, those two combinations (with projectors
given by (\ref{pro})) leave the supersymmetry unbroken. This shows
that the F-term string in the Bogomol'nyi bound is $\half$-BPS
saturated.

The semilocal extension of the F-term model is obvious with a second
pair of charged superfields $\Psi_+,\Psi_-$ and superpotential \be
W=\beta\Phi_0(\Phi_+\Phi_- + \Psi_+\Psi_- - {\eta^2 \over 2}) \ee
However in this case the scalar potential is (note that $\psi_\pm$ are again scalars)
\be
U=\beta^2\left|\phi_+\phi_-+\psi_+\psi_--\frac{\eta^2}{2}\right|^2+\beta^2|\phi_0|^2\left(|\phi_+|^2+|\phi_-|^2+|\psi_+|^2+|\psi_-|^2\right)
+\frac{g^2}{8}\left(|\phi_+|^2+|\psi_+|^2-|\phi_-|^2-|\psi_-|^2\right)^2
\ee and it can be away from the Bogomol'nyi limit, so the stability of
the string solutions depends on the values of the couplings. If $\beta
< g^2/2$ the Nielsen-Olesen strings are stable but if $\beta > g^2 /2$
no strings will form (see also \cite{EY03}).

In the Bogomol'nyi limit the configurations that saturate the bound
have $\phi_0= 0,\ \phi_+=\phi_-^*, \ \psi_+=\psi_-^* $. Again there
is no unique string solution, but a one-parameter family of
string-like configurations where the magnetic flux is spread over an
arbitrarily large area. They can be obtained from the D-term ones by
an SU(2) rotation described in the next section.

\section{The equivalence between D-term and F-term models: P-term models}

We have given arguments suggesting that the F-term model in the
Bogomol'nyi limit and the D-term model in the super-Bogomol'nyi limit
are both $N\=2$ and admit string solutions that break half of the
supersymmetries. Both statements can be proved easily in the $N\=2$
formalism. In fact, from the $N\=2$ point of view, the two models are
equivalent \cite{DKP04} and we will now illustrate this equivalence
(once again, full justification relies on the $N\=2$ analysis).

As anticipated, in $N\=2$ language the neutral scalar field and the
gauge field belong to the same (``vector'') multiplet. The auxiliary
field of the vector multiplet is an $SU(2)$ triplet $\vec P$ called a
P-term in this case, and from the $N\=1$ point of view it reduces to F-
or D-terms for different orientations. Therefore, the $SU(2)$ symmetry
existing among the $N\=2$ supercharges rotates F-terms into D-terms.
Charged superfields also combine in a non-chiral hypermultiplet of
charge $q = +1$, $h = (h_1,h_2)=(\phi_+, \phi_-^*)$ ($h_i$ where
$i=1,2$ are in the fundamental representation of $SU(2)$).

One can add Fayet-Iliopoulos terms of the form $\vec \xi \cdot \vec P$.
The resulting Lagrangian, after elimination of the auxiliary fields,
has the following potential term for the scalars:

\be U= {g^2\over 8} \sum_{i=1}^3 (h^\dagger\tau^i h - \xi^i)^2\ee that
is, \be U = {g^2\over 8} \bigl[ (\phi_+^*, \phi_-) \tau^1
\pmatrix{\phi_+ \cr \phi_-^*\cr} - \xi^1\bigr]^2 + {g^2\over 8} \bigl[
(\phi_+^*, \phi_-) \tau^2 \pmatrix{\phi_+ \cr \phi_-^*\cr} -
\xi^2\bigr]^2 + {g^2\over 8} \bigl[ (\phi_+^*, \phi_-) \tau^3
\pmatrix{\phi_+ \cr \phi_-^*\cr} - \xi^3 \bigr]^2 \ee where $\tau^i$
are the Pauli matrices. The D-term model is obviously obtained if
$\xi_1=\xi_2=0$. To see the relation to F-terms, note that
$$\biggl|\phi_+\phi_- -{\eta^2\over 2}\biggr|^2 = \biggl[{\rm Re} (\phi_+\phi_-
-{\eta^2\over 2})\biggr]^2 + \biggl[{\rm Im} (\phi_+\phi_-)\biggr]^2 = {1\over 4} \biggl[
(\phi_+^*, \phi_-) \tau^1 \pmatrix{\phi_+ \cr \phi_-^*\cr} -
{\eta^2}\biggr]^2 + {1\over 4}\biggl[ (\phi_+^*, \phi_-) \tau^2
\pmatrix{\phi_+ \cr \phi_-^*\cr}\biggr]^2
$$ so it corresponds to $\xi_2=\xi_3=0$. 

An SU(2) rotation $h_i \to
U_i^j h_j$ \be \pmatrix{\phi_+ \cr \phi_-^*\cr} \to {1\over \sqrt 2}
\pmatrix{ \cos{\theta \over 2} + i\sin{\theta\over 2}n^3
&i\sin{\theta\over 2} (n^1-in^2)\cr i\sin{\theta\over 2} (n^1+in^2)
&\cos{\theta \over 2} - i\sin{\theta\over 2}n^3 \cr} \pmatrix{\phi_+
\cr \phi_-^*\cr} \ee
will induce an (inverse) SO(3) rotation on the $\xi^i$ of angle
$-\theta$ about the $ \vec{n}$ axis ($\vec{n} \cdot \vec{n} = 1$).
For instance, to go from D-term $\vec \xi = (0,0, \eta^2)$ to F-term
$\vec \xi = (\eta^2,0,0)$ we need a rotation of $-\pi/2$ about the
$y$-axis, which is obtained by an SU(2) transformation\footnote{This
transformation differs from the analogous one in \cite{KL03} by a
factor of $i\sigma^2$} of the scalars with $\theta = \pi/4$, $\vec{n}
= (0,1,0)$: \be\pmatrix{\phi_+ \cr \phi_-^*\cr} \to {1\over \sqrt 2}
\pmatrix{1 &1\cr -1&1\cr} \pmatrix{\phi_+ \cr \phi_-^*\cr} \ee in
agreement with what we found in section \ref{Ftermstrings}.

\section{Discussion}

Let us summarise here the main points illustrated in this paper:

$\bullet$ D-term models are always in the Bogomol'nyi limit, but can be
$N\=1$ or $N\=2$ depending on the superpotential. In all cases the
Nielsen-Olesen string breaks half of the supersymmetries, so in the
superBogomol'nyi limit the number of fermion zero modes is twice that
outside the Bogomol'nyi limit. Outside the superBogomol'nyi limit there
are two chiral zero modes, and a second pair of antichiral zero modes
appear in the superBogomol'nyi limit.

$\bullet$ F-term strings are $N\=1$ supersymmetric even away from the
Bogomol'nyi limit, and become $N\=2$ in the Bogomol'nyi
limit. Supersymmetry breaking is complete in the $N\=1$ case and there is
partial breaking in the $N\=2$ case. As a result, the number of fermion
zero modes is always four, two of each chirality.

$\bullet$ The F-term model in the Bogomol'nyi limit is equivalent to
the D-term model in the superBogomol'nyi limit. This is obvious in the
manifestly $N\=2$ supersymmetric formulation, where the two are related
by an $SU(2)$ rotation of the supercharges.

$\bullet$ D-term models with zero superpotential have a flat direction
$|\phi_+|^2 - |\phi_-|^2 = \eta^2$. The Bogomol'nyi equations force
$\phi_-=0$ for the BPS string (a ``vacuum selection''
effect). Addition of a superpotential $W =\beta \Phi_0\Phi_+\Phi_-$
lifts the flat direction but does not change the structure of the
D-term strings. For a specific value of $\beta$ a second supersymmetry
appears of the opposite chirality.

$\bullet$ Both types of models can be made semilocal by the addition
of a second pair (or more) of charged superfields with identical
couplings. The stability of strings in the semilocal model depends on
the ratio of the scalar and gauge masses, so F-term models can only
have stable strings if $\lambda < e^2$ (the same condition as for
type-I superconductors) or neutrally stable strings in the Bogomol'nyi
limit. D-term models are always in the Bogomol'nyi limit so they lead
to neutrally stable strings.

This raises a number of interesting open questions:

- In D-term models the semilocal zero mode survives coupling
to gravity, including $N\=1$ supergravity, and this is important for the
viability of D-term inflation, and in particular of some brane
inflation models \cite{UAD04,BDKP04,DHKLZ04,H04,H04a,H04b}. But it also
shows that the Nielsen-Olesen D-term strings are not the only
string-like BPS states in $N\=1$ supergravity, and the task of
identifying D-strings in the low energy limit may not be as
straightforward as had been hoped.

- In the F-term case the situation is unclear, and the fate of
the zero mode has not been analysed so far. We expect the answer to
depend on the details of the supersymmetry breaking \cite{PU03,EY03}.

Let us get back to the paradox mentioned in the introduction. If one
looks at the bosonic sector alone, both F-term strings and D-term
strings satisfy Bogomol'nyi bounds which are known to survive coupling
to (bosonic) gravity \cite{CG87}.  In fact we have shown that before
coupling to gravity, the D-term and F-term models considered in
\cite{DKP04} are $N\=2$ supersymmetric and their D-term and F-term
strings are completely equivalent. They both carry a conserved
topological charge, which appears as a central charge (current) in the
supersymmetry algebra, and the naive expectation is that they should
both survive as BPS states in the presence of gravity.  And yet, it
was shown in \cite{DKP04} that in the $N\=1$ supergravity model only the
D-term strings remain BPS; F-term strings break all the
supersymmetries {\it even in the Bogomol'nyi limit}.

The problem is that coupling to $N\=1$ supergravity involves making local
{\it one} of the two supersymmetries present in the Bogomol'nyi limit,
and we have seen that the broken and unbroken supersymmetries are
specific combinations. If we make local one of the combinations
${(Q_1\pm Q_2) /\sqrt 2}$, the resulting $N\=1$ strings will appear to
be D-term strings and exhibit partial breaking of (now local)
supersymmetry as described in \cite{DKP04}. Any other choice, for
instance making $Q_1$ local, gives rise to a complete breaking of the
local supersymmetry by the strings. This is the choice that was
implicit in \cite{DKP04}. In terms of the central charges in the
supersymmetry algebra, the difference is that in the F-term model the
central charge appears in the mixed commutators of the two
supersymmetries, so BPS states are only possible in the $N\=2$ case,
while in the D-term model the central charge appears in each of the
two $N\=1$ subalgebras, so partial breaking of supersymmetry is possible
in both the $N\=1$ and $N\=2$ cases \cite{GS00}.

\bigskip
We would like to end with a comment about the cosmological relevance
of BPS solutions and partial breaking of supersymmetry. Dvali, Kallosh
and van Proeyen were interested in identifying how certain superstring
and brane configurations look like from the low energy
(i.e. supergravity) point of view. However, there is another reason
why partially broken supersymmetry is important for cosmology: it can
give rise to chiral vortons, whose cosmological implications can be
quite serious as they can disrupt nucleosynthesis or even overclose the
mass of the Universe unless they are formed at a relatively low
energy\cite{CP99,CD00,DKPS00,PD01}. This was the original motivation
behind the study by Davis, Davis and Trodden, and it is still
valid. The presence of chiral vortons imposes serious constraints on
any particle physics model, so in principle it could also help to
constrain superstring/M theory compactifications.

\section{acknowledgments}
We are grateful to the authors of refs. \cite{DDT97,DKP04} for
discussions, especially Anne Davis and Toine van Proeyen, and to Alexei Yung
for pointing out references \cite{VY01,EY03}. We also thank
Eric Bergshoeff, Jan de Boer, Mboyo Esole, Filipe Freire, Mees de
Roo, Petja Salmi and Stefan Vandoren. J.U. thanks the University of Leiden, where part of this work was done, for hospitality. This work was supported by the
ESF Programme COSLAB - Laboratory Cosmology, the Netherlands
Organization for Scientific Research (NWO) under the VICI programme
and FPA 2002-02037 and 9/UPV00172.310-14497/2002.
J.U. is supported by the Spanish {\it Secretar\'\i a de Estado de
Educaci\'on y Universidades} and {\it Fondo Social Europeo}. 

\bigskip\hrule\bigskip

\bibliography{biblio.bib}

\end{document}